\documentclass[twocolumn]{aastex63}

\usepackage[caption=false]{subfig}
\usepackage{rotating}
\usepackage{color, colortbl}
\usepackage{tikz}
\usepackage{afterpage}
\usepackage[para]{threeparttable}
\usepackage{tabularx}
\usepackage{graphicx}

\usepackage{lineno}

\shorttitle{Bias-corrected RSG luminosity function}
\shortauthors{Strotjohann et al.}
\graphicspath{{./}{figures/}}

\accepted{by ApJ Letters on February 19, 2024}

\begin{document}

\title{A bias-corrected luminosity function for red supergiant supernova progenitor stars}

\author[0000-0002-4667-6730]{Nora L. Strotjohann}
\affiliation{Department of Particle Physics and Astrophysics, Weizmann Institute of Science, 76100 Rehovot, Israel}

\author{Eran O. Ofek}
\affiliation{Department of Particle Physics and Astrophysics, Weizmann Institute of Science, 76100 Rehovot, Israel}

\author{Avishay Gal-Yam}
\affiliation{Department of Particle Physics and Astrophysics, Weizmann Institute of Science, 76100 Rehovot, Israel}

Accepted by ApJ Letters

\begin{abstract}
The apparent tension between the luminosity functions of red supergiant (RSG) stars and of RSG progenitors of Type II supernovae (SNe) is often referred to as the RSG problem and it motivated some to suggest that many RSGs end their life without a SN explosion. However, the luminosity functions of RSG SN progenitors presented so far were biased to high luminosities, because the sensitivity of the search was not considered. Here, we use limiting magnitudes to calculate a bias-corrected RSG progenitor luminosity function. We find that only $(36\pm11)\%$ of all RSG progenitors are brighter than a bolometric magnitude of $-7\,\text{mag}$, a significantly smaller fraction than $(56\pm5)\%$ quoted by \citet{davies2020}. The larger uncertainty is due to the relatively small progenitor sample, while uncertainties on measured quantities such as magnitudes, bolometric corrections, extinction, or SN distances, only have a minor impact, as long as they fluctuate randomly for different objects in the sample.
The bias-corrected luminosity functions of RSG SN progenitors and Type M supergiants in the Large Magellanic cloud are consistent with each other, as also found by \citet{davies2020} for the uncorrected luminosity function. The RSG progenitor luminosity function, hence, does not imply the existence of failed SNe.
The presented statistical method is not limited to progenitor searches, but applies to any situation in which a measurement is done for a sample of detected objects, but the probed quantity or property can only be determined for part of the sample.
\end{abstract}


\section{Introduction} \label{sec:intro}

The red supergiant (RSG) problem was first proposed by \citet{li2006_rsg_problem}, \citet{kochanek2008_failedsne}, and \citet{smartt2009} who state that detected RSG supernova (SN) progenitors are fainter than expected according to stellar evolution models. This observation is sometimes interpreted as evidence that luminous RSGs with initial masses of $\gtrsim 17\,\text{M}_\odot$ collapse into black holes directly without producing a SN-like optical transient (see, e.g., \citealt{smartt2015,kochanek2020}). Indeed, such failed SNe are predicted by nuclear calculations (see, e.g., \citealt{patton2020, burrow2020_failed_sne, boccioli_2023_failed_sne} for recent simulations).

However, the significance of the RSG problem has come under debate: \citet{davies2018} argue that SN progenitors have lower surface temperatures than assumed by \citet{smartt2009} and \citet{smartt2015}. The resulting, larger bolometric corrections boost the inferred progenitor luminosities and masses. Another difference is that \citet{davies2020} compare the progenitor luminosity function to the luminosities of detected RSGs in the Large Magellanic cloud (LMC) and do not rely on stellar evolution models. They conclude that the RSG problem is less significant than $2\,\sigma$.

The issue of bolometric corrections is entangled with the typically unknown amount of extinction (see \citealt{maund2017_progenitor_stellar_environment} for host extinction estimates). Additional extinction can arise from shells of circumstellar medium (CSM) or dust around the progenitor. But depending on the geometry and dust properties such shells can also scatter light into the line of sight and reemit absorbed radiation. These effects can reduce the impact of dust in some bands and redden the observed flux (see, e.g., \citealt{kochanek2012} and \citealt{kilpatrick2023_sn2023ixf_progenitor}). CSM properties and absorption might even change within the last years before the SN (see calculations by \citealt{davies2022}) as the impending SN core collapse might trigger eruptive mass-loss events on the surface of some progenitor stars (see, e.g., \citealt{strotjohann2021,jacobson-galan2022}).

Another difficulty is that the progenitor stars of some SNe might not have been identified correctly (see, e.g., \citealt{maund2015}). Many SNe have not yet faded sufficiently to confirm the disappearance of the progenitor and some progenitors are located in unresolved stellar clusters or have binary companions. Moreover, the uncertainties on the distances of nearby galaxies are large. Both effects are important because the sample of detected progenitors is relatively small, such that discarding a few objects or revising their luminosities can have a major impact on the luminosity function.

In addition to the complications described so far, here, we point out a statistical bias: In previously presented luminosity functions the sensitivity of progenitor searches was not considered correctly. As a consequence, the fraction of intermediate-luminosity progenitors was overestimated resulting in a steeper, cumulative luminosity function that appears to have a cutoff at high luminosities. We correct for this specific bias and show that the shape of the resulting RSG SN progenitor luminosity function is consistent with the luminosities of Type M supergiants in the LMC. We also quantify the uncertainty due to the small sample size (previously done for the brightest progenitor by \citealt{davies2018}) and find that it dominates the uncertainty on the luminosity function. In comparison, uncertainties on the progenitor bolometric magnitudes only have a minor impact as long as they fluctuate randomly for different progenitors and are not systematically biased.

In Sect.~\ref{sec:fraction} of this paper, we calculate the fraction of progenitor stars that are brighter than a given magnitude threshold while correcting for the sensitivity of the search (Sect.~\ref{sec:bernoulli}). A binomial confidence interval quantifies the uncertainties due to statistical fluctuations (Sect.~\ref{sec:confidence_intervals}) and we consider in addition measurement uncertainties on extinction, bolometric corrections, SN distances, progenitor magnitudes, and limiting magnitudes (Sect.~\ref{sec:lf_with_errors}). In Sect.~\ref{sec:lf_unbiased}, we present the bias-corrected luminosity function and highlight how it differs from previously published luminosity functions. In Sect.~\ref{sec:rsg_problem}, we compare the luminosity functions of RSG SN progenitors and detected RSGs in the LMC and conclude in Sect.~\ref{sec:conclusion}.

\section{Constraining the fraction of bright RSG progenitors} \label{sec:fraction}

In this section we consider a single luminosity threshold and estimate the fraction of brighter RSG progenitors and the uncertainty on the fraction. The cumulative luminosity function is obtained in Sect.~\ref{sec:lf_unbiased} by repeating the calculation for many luminosity steps.

\subsection{Considering the search sensitivity}
\label{sec:bernoulli}

We compile RSG SN progenitor searches from the literature (see Table~\ref{tab:rsg_progenitors}) and list the progenitor magnitudes or limiting magnitudes obtained in the original single-band observations. Next, we correct these magnitudes for extinction, apply the bolometric corrections by \citet{davies2018}, and subtract the distance modulus. The resulting absolute bolometric magnitudes and limits are given in the second- and third-to-last columns of Table~\ref{tab:rsg_progenitors}.

To calculate a bias-corrected luminosity function, we have to account for the sensitivity of each search. We use limiting magnitudes to quantify down to which magnitude progenitor stars would still be detectable. We consider the detection probability a step function that jumps from one to zero at the limiting magnitude. In practice, this transition is less abrupt.

Limiting magnitudes are typically not quoted if the progenitor star is detected. In those cases, we estimate them by assuming that the error on the progenitor magnitude is dominated by the noise level of the background (similar to \citealt{kochanek2020}) and we convert the errors to $3\,\sigma$ limiting magnitudes (given in the sixth column of Table~\ref{tab:rsg_progenitors}). Additional error sources might contribute and, therefore, it would be preferable to reanalyze the observations and measure limiting magnitudes from the images directly.

\begin{figure}[t]
\centering
\includegraphics[width=\columnwidth]{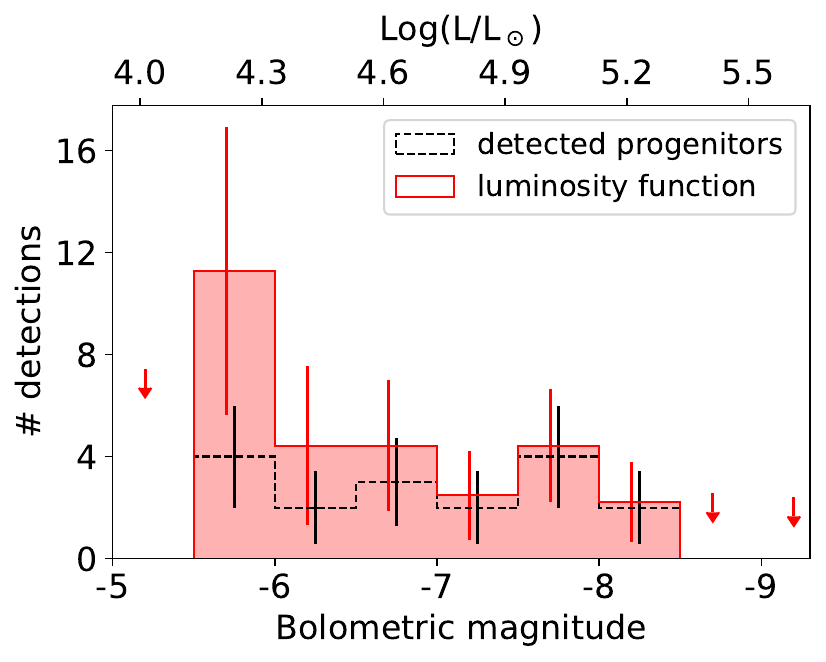}
\\
\caption{\label{fig:lf_pdf} The non-cumulative, bias-corrected RSG SN progenitor luminosity function. The black distribution shows the number of detected RSG progenitors taken from Table~\ref{tab:rsg_progenitors} with $1\,\sigma$ Poisson errors. We then divide by the fraction of sensitive searches in each bin. This correction is small for bright progenitors, but only 11 out of 31 searches are sensitive enough to detect progenitors fainter than an absolute bolometric magnitude of $-6\,\text{mag}$. While four as faint progenitors were detected, we would have expected $11.3$ detections if all searches had been as sensitive and the red distribution is scaled up accordingly. Arrows mark 90\% confidence limits for bins without progenitor detections.}
\end{figure}

Figure~\ref{fig:lf_pdf} shows how limiting magnitudes can be used to calculate a bias-corrected, non-cumulative luminosity function. In the following, we consider cumulative luminosity functions instead and quantify the uncertainties more carefully, but the underlying principle remains the same.

\begin{figure*}[t]
\centering
\subfloat[Binomial likelihood function and $1\,\sigma$ confidence intervals based on different prescriptions]{\includegraphics[width=0.99\columnwidth]{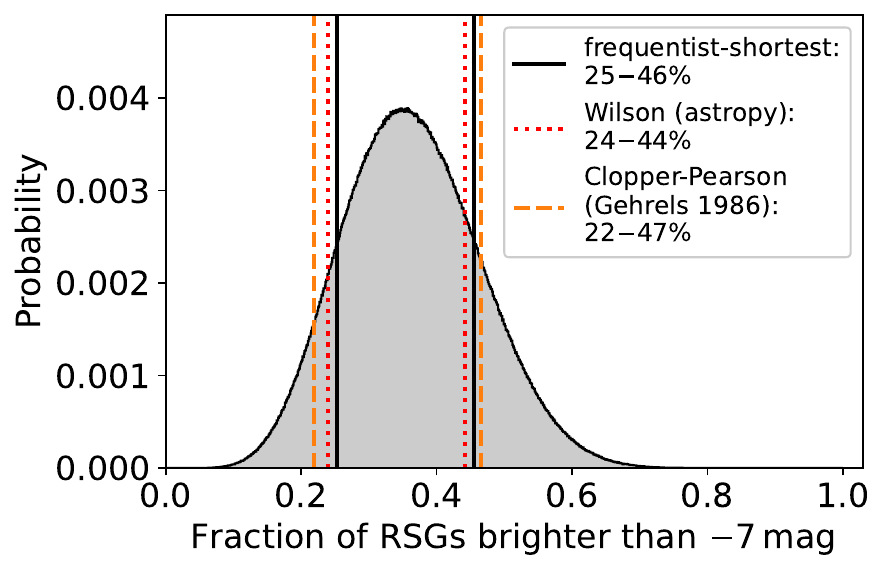}\label{fig:conf_intervals} } \hfill
\subfloat[Likelihood function including measurement uncertainties]{\includegraphics[width=0.99\columnwidth]{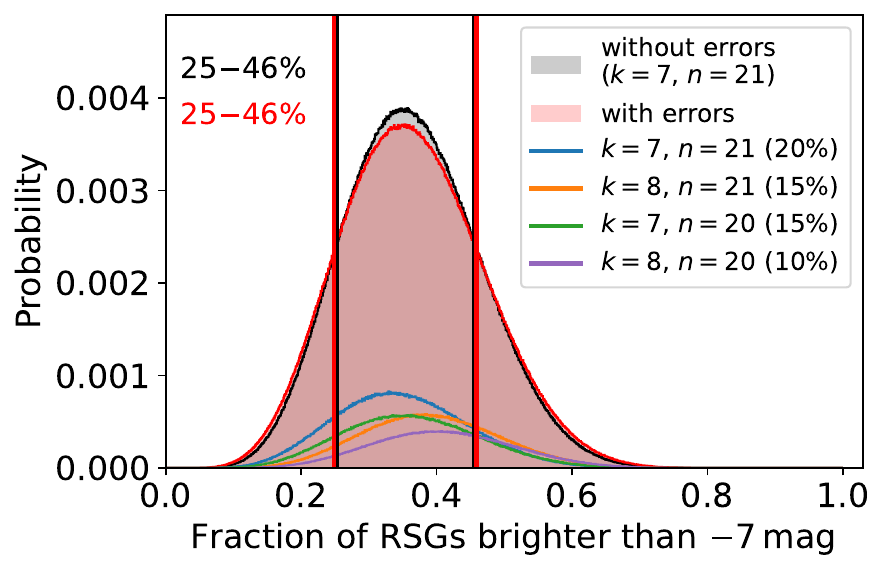}\label{fig:conf_with_errs} }
\caption{\label{fig:confidence_calc} Left: Likelihood function for the fraction of RSG progenitors brighter than $-7\,\text{mag}$, based on 7 detections in 21 sensitive searches. The vertical lines show different $1\,\sigma$ confidence intervals (see Sect.~\ref{sec:confidence_intervals}) and we use the frequentist-shortest interval in the following. Right: Likelihood function without (black) and with (red) measurement uncertainties. The vertical lines indicate the $1\,\sigma$ confidence regions for the two distributions. When allowing progenitor magnitudes and limiting magnitudes to vary within their uncertainties, the number of sensitive searches is not always 21, but varies between 17 and 24 searches with 4 to 10 detections. Accordingly, we obtain many, slightly different distributions and the colored curves show the four most frequent ones. When summing over all distributions, we obtain the red likelihood function that peaks at $\hat{p}=36\%$ with a $1\,\sigma$ region of $25-46\%$. Neglecting measurement errors would yield the gray distribution instead which is only marginally broader in this case.}
\end{figure*}

To estimate the (cumulative) fraction of bright progenitors, $\hat{p}(t)$, we adopt a magnitude threshold, $t$, and only select observations with limiting magnitudes deeper than the threshold. This ensures that we consider searches if and only if they are sufficiently sensitive regardless of whether they yield detections. The fraction of brighter progenitors is then calculated only among these searches and is given by
\begin{equation}
\hat{p}(t) = \frac{k(t)}{n(t)} = \frac{\text{\# searches with limmag} > t \text{ and mag} < t}{\text{\# searches with limmag} >t }
\end{equation}
For illustration, the second-to-last column of Table~\ref{tab:rsg_progenitors} shows that $n=21$ out of the 31 searches are sensitive to progenitors that are brighter than a bolometric magnitude of $t=-7\,\text{mag}$. $k=7$ out of 21 progenitors are brighter than the chosen magnitude threshold and we hence measure a fraction of $\hat{p}(t=-7\,\text{mag}) = 33\%$ bright progenitors for this example.

\subsection{Calculating Binomial confidence intervals}
\label{sec:confidence_intervals}

When fixing the magnitude threshold, $t$, progenitor searches can be considered \emph{Bernoulli} experiments. For an unknown true fraction of bright progenitors, $p$, and $n$ sensitive searches, the Binomial distribution describes how likely it is to detect exactly $k$ progenitors:
\begin{equation}
\label{eq:binomial}
    P(k,n,p) = \frac{n!}{k! (n-k)!} p^k (1-p)^{n-k} \quad .
\end{equation}
The most likely estimate for the fraction of bright progenitors is $\hat{p} = k/n$ (see Sect.~\ref{sec:bernoulli}). To calculate an uncertainty on $\hat{p}$, we construct the \citet{neyman1937} confidence interval: We let $p$, the true fraction of bright progenitors, vary between 0 and 1 and set $n$ and $k$ to the number of sensitive searches and detected progenitors, respectively. The gray distribution in Fig.~\ref{fig:conf_intervals} shows the resulting likelihood function for a magnitude threshold of $t=-7\,\text{mag}$, corresponding to $n=21$ sensitive searches (see Table~\ref{tab:rsg_progenitors}) and $k=7$ bright progenitors.

To construct a confidence interval, we select the part of the likelihood function in Fig.~\ref{fig:conf_intervals} that contains the desired fraction of the distribution, e.g., 68\% for a $1\sigma$ confidence interval.
However, this interval can be defined in several ways, because we cannot match the desired coverage for every value of $p$ due to the discrete nature of the Binomial distribution (Eq.~\ref{eq:binomial}; see, e.g., \citealt{cousins2010_review_binomial_ci} for a review).

Confidence limits calculated with the \citet{clopper_pearson1934} method (also used by \citealt{gehrels1986_confidencelimits}) are shown by orange dashed lines in Fig.~\ref{fig:conf_intervals}. This method yields a central confidence interval and guarantees overcoverage for all values of $p$, but it is overly conservative for small sample sizes with $n \lesssim 100$ (see e.g. \citealt{brown2001_binomial_ci, cousins2010_review_binomial_ci, cameron2011_binomial_confinterval, thulin2013_binomialci}). Here, for 21 searches, the $1\sigma$ confidence interval has an average coverage of $80.1\%$ instead of $68.3\%$ (see e.g. \citealt{cameron2011_binomial_confinterval}). 
An alternative is the approximate \citet{wilson1927} score interval implemented in \emph{astropy} \citep{astropy:2013, astropy:2018, astropy:2022} or the beta distribution recommended by \citet{cameron2011_binomial_confinterval}. Both methods yield central intervals and the coverage is correct when averaging over the possible values of $p$, but not for every $p$. The deviations are especially large, when $p$ is close to 0 or 1. The Wilson score interval is shown by red, dotted lines in Fig.~\ref{fig:conf_intervals}.

Here, we decide to use the non-central, frequentist-shortest confidence interval. We calculate it numerically by excluding the bins with the smallest probabilities in Fig.~\ref{fig:confidence_calc} until the remaining bins match the desired average coverage. The advantages are that we obtain the shortest possible confidence interval, the prescription switches naturally between one- and two-sided intervals, and it is not limited to the Binomial probability, but can be calculated numerically for any likelihood shape. The resulting interval is indicated by vertical, black lines in Fig.~\ref{fig:conf_intervals} and is very similar to the Wilson approximation.

Note, that we consider the number of sensitive searches a fixed quantity, i.e., we calculate the fraction of bright progenitors given the observations (see \citealt{cousins2006_literature_review}). This is called \emph{conditioning}. Alternatively, one could assume that the number of sensitive searches $n$ is drawn from a Poisson distribution with an unknown expectation value (see e.g. \citealt{cousins2010_review_binomial_ci}). This would shift the $1\,\sigma$ confidence region in Fig.~\ref{fig:conf_intervals} to $27 - 46\%$ with $\hat{p}=35\%$.

\subsection{Including measurement errors}
\label{sec:lf_with_errors}

Next, we quantify the impact of measurement errors on the luminosity function by generating many pseudo-experiments and allowing the quantities to fluctuate within their uncertainties listed in Table~\ref{tab:rsg_progenitors}. Following \citet{davies2020}, we assume that the errors on the SN distance, magnitudes, and extinction follow a normal distribution, but set negative extinction values to zero. We draw bolometric corrections from a uniform distribution within the limits quoted by \citet{davies2018}. For limiting magnitudes, we adopt an error of $0.1\,\text{mag}$, corresponding to a $\sim10\%$ uncertainty on the noise level. Importantly, we assume that the errors are random and fluctuate independently for the different progenitors in the sample. Bolometric corrections, SN distances, or extinction values could also be systematically offset for all or many progenitors. If known, or suspected, such shifts can be considered similarly by assigning the simulated value to the entire sample instead of drawing a random value for each object separately.

An easy and intuitive way of considering both statistical and systematic errors is absorbing them into the likelihood function. This approach is the standard error treatment in Bayesian statistics, but it is sometimes adapted in frequentist statistics (see, e.g., \citealt{cousins1992_systematic_uncertainties, conrad2003_nuisanceparameters_in_pdf}) like in our case.

For the different pseudo-experiments, the number of sensitive searches, $n$, and detections, $k$, at the magnitude threshold can vary. For a magnitude threshold of $-7\,\text{mag}$, considered in the example above, the colored lines in Fig.~\ref{fig:conf_with_errs} show the most common combinations of $n$ and $k$: We still obtain 7 detections and 21 sensitive searches for 20\% of the experiments, however, in 15\% of the cases, we detect 8 progenitors instead and in an other 15\% only 20\% of the searches are more sensitive than $-7\,\text{mag}$. Summing over the tens of different combinations of $n$ and $k$ yields the red distribution in Fig.~\ref{fig:conf_with_errs}. For the chosen threshold of $-7\,\text{mag}$, including errors barely makes a difference, but Fig.~\ref{fig:lf_bolo_witherrs} shows the distribution is shifted and broadened for other magnitude thresholds. This happens when one or few progenitor stars that have detections or limiting magnitudes close to the chosen threshold.

Measurement errors can, hence, be included by resampling progenitor magnitudes and limits many times while allowing them to vary within the estimated uncertainties. However, for the considered example, measurement errors only play a minor role. Instead, the uncertainty on the fraction of bright progenitors is dominated by statistical fluctuations due to the relatively small sample size (see Sect.~\ref{sec:confidence_intervals}).

\section{Comparing to previously presented luminosity functions} \label{sec:lf_unbiased}

\begin{figure}[tb]
\centering
\includegraphics[width=\columnwidth]{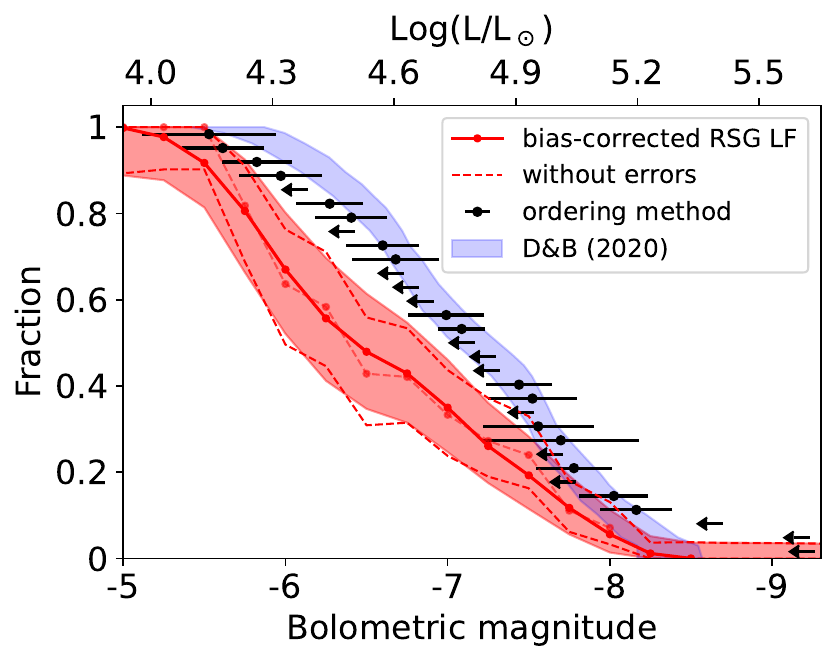}
\\
\caption{\label{fig:lf_bolo_compare_methods} The RSG progenitor luminosity function (red) with its $1\,\sigma$ confidence region due to statistical fluctuations and measurement uncertainties on the progenitor bolometric magnitudes. Neglecting measurement errors would yield the red, dashed distribution instead. Previously published luminosity functions, simply ordered detections and upper limits by magnitude and the resulting distribution is shown by black data points. The $1\,\sigma$ confidence region of the \citet{davies2020} luminosity function is shown in blue. They largely follow the ordering method, but additionally quantify the impact of measurement errors. Deviations from the ordering method are mostly caused by the updated RSG progenitor sample.}
\end{figure}

Next, we construct the entire luminosity function, by repeating the calculation described in Sect.~\ref{sec:fraction} for bolometric magnitudes between $-5\,\text{mag}$ and $-10\,\text{mag}$. The red band in Fig.~\ref{fig:lf_bolo_compare_methods} shows the bias-corrected RSG luminosity function and its 68\% uncertainty due to Binomial fluctuations as well as measurement uncertainties. For comparison, the red dashed lines show the luminosity function without measurement errors. It has edges because the number of sensitive searches and detections can change abruptly between bins. Considering measurement errors, hence, smoothes out the luminosity function, but only increases the size of the confidence region marginally.

In previously presented RSG luminosity functions, authors do not consider the sensitivity of progenitor searches, but sort both detections and upper limits by magnitude to construct a cumulative luminosity function (see e.g., \citealt{smartt2009,smartt2015,davies2018, davies2020,kochanek2020,davies2020b}). We call this approach the \emph{ordering method} and the resulting luminosity function is shown by black data points in Fig.~\ref{fig:lf_bolo_compare_methods}.

The ordering method effectively treats upper limits like detections and therefore biases the function to a higher fraction of bright progenitors. It would yield the correct result only if every search resulted in a detection, i.e., the sensitivity can only be neglected if the entire target source population is brighter than the limiting magnitude of the search. 
This condition is fullfilled for complete, flux-limited samples, such as the SN sample collected by the Bright Transient Survey \citep{fremling2020, perley2020} or the \emph{Swift} GRB sample \citep{pescalli2016_grb_lf}. However, in the presence of non-detections, the ordering method provides an upper limit on the true, cumulative luminosity function, but does not recover its shape.

\citet{davies2020} calculate the luminosity function in a manner similar to the ordering method, but additionally quantify the impact of measurement errors using a Monte Carlo simulation that we adapted in Sect.~\ref{sec:lf_with_errors}. In the Monte Carlo simulation, they also allow undetected progenitors to take on lower luminosities. However, this requires choosing a lower-luminosity threshold and the method leads to biases if the true luminosity function is not flat in the relevant magnitude range. For example, including several rather unconstraining non-detections into the sample would induce a bias to bright progenitors. Figure~4 of \citet{davies2020} shows that the resulting luminosity function is nearly identical to the one obtained with the ordering method. We show their 68\% confidence region in blue in Fig.~\ref{fig:lf_bolo_compare_methods}. The deviations from the ordering method are here mostly due to the updated RSG sample.

Figure~\ref{fig:lf_bolo_compare_methods} shows that correcting for the sensitivity yields a significantly lower cumulative luminosity function. It is less steep at intermediate magnitudes between $-7.5\,\text{mag}$ and $-6.5\,\text{mag}$, but requires a larger fraction of RSGs that are fainter than $-6.5\,\text{mag}$. Most progenitor searches are sensitive to the brightest progenitors with bolometric magnitude of $-8\,\text{mag}$, therefore, neglecting the sensitivity does not have a dramatic impact on the high-luminosity end of the luminosity function.

Another advantage of the bias-corrected luminosity function is that statistical errors due to the sample size are included naturally (see Sect.~\ref{sec:confidence_intervals}).
The red band in Fig.~\ref{fig:lf_bolo_compare_methods} is much wider than the measurement errors calculated by \citet{davies2020} (blue band). The uncertainty on the luminosity function is hence currently dominated by the relatively small sample size and uncertainties on the measured magnitudes, bolometric corrections, extinction, and SN distances are small as long as they are not systematically biased.

The faintest detected progenitor has a bolometric magnitude of $-5.5\,\text{mag}$ and the derived luminosity function (shown in Fig.~\ref{fig:lf_bolo_compare_methods}) requires that $<12\%$ of the progenitors are fainter at $1\sigma$ confidence level.  
No detected RSG progenitors are brighter than $-8.25\,\text{mag}$ and the $1\sigma$ upper limit on the fraction of brighter RSG progenitors is $<3.4\%$.

\section{The RSG problem} \label{sec:rsg_problem}

\begin{figure}[t]
\centering
\includegraphics[width=\columnwidth]{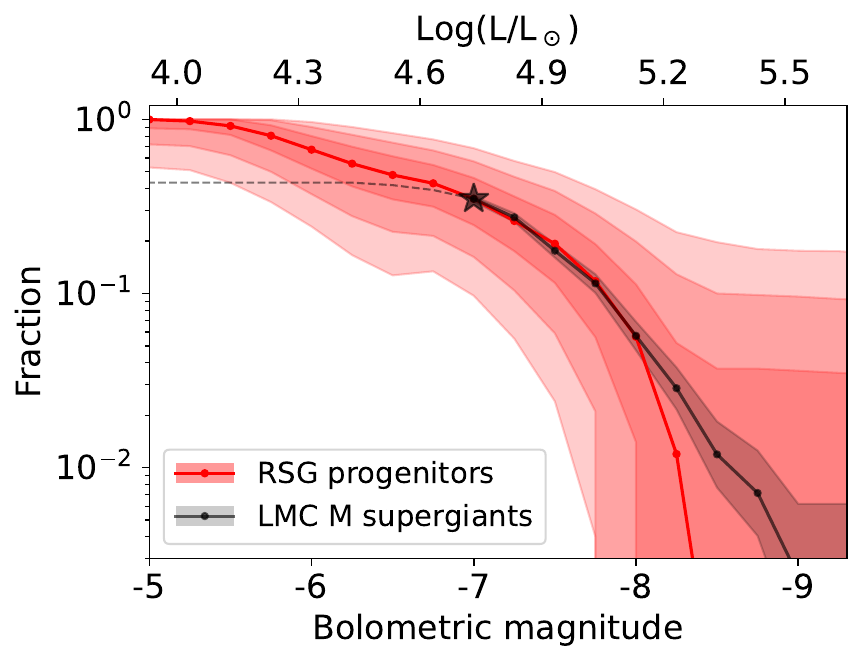}
\\
\caption{\label{fig:lf_bolo_witherrs} The cumulative luminosity function of RSG SN progenitor stars (also shown in Fig.~\ref{fig:lf_bolo_compare_methods}) and its 1, 2, and $3\,\sigma$ confidence regions in red compared to the luminosity function of M supergiants in the LMC (in gray; sample from \citealt{davies2018b}). The gray star indicates that the LMC is normalized to the progenitor sample at $-7\,\text{mag}$, because it is incomplete at fainter magnitudes. The two distributions are fully consistent and do not provide any evidence for an RSG problem. The bins of a cumulative luminosity function are not independent of each other, therefore, the scatter within the distribution seems small compared to the shown uncertainty. The confidence region indicates by how much the distribution might shift up or down when increasing the sample size.}
\end{figure}

To quantify whether the RSG problem is significant, we compare in Fig.~\ref{fig:lf_bolo_witherrs} the measured, cumulative luminosity function for RSG SN progenitor stars (red) to the luminosity function of M supergiants in the LMC (gray; sample from \citealt{davies2018b}). The red-shaded bands indicate the $1$, $2$, and $3\sigma$ confidence regions and we chose a logarithmic y-axis to highlight the high-luminosity end of the luminosity function.

The LMC sample is complete for bolometric luminosities $>10^{4.7}\text{L}_\odot$ \citep{davies2018b}, corresponding to a bolometric magnitude of $<-6.9\,\text{mag}$. Therefore, we normalize the LMC sample to the progenitor luminosity distribution at $-7\,\text{mag}$ and only compare the two distributions at higher luminosities. The normalized LMC M supergiant luminosity function is shown as a gray line in Fig.~\ref{fig:lf_bolo_witherrs} and the gray-shaded region shows its $1\sigma$ uncertainty due to the sample size. 
For fainter magnitudes, we show the LMC luminosity function as a dashed line, because it is likely not complete: While only few detected LMC RSGs are fainter than $-6.5\,\text{mag}$, $>50\%$ of the RSG progenitors are as faint. This would imply that we only detect the brighter half of the RSGs in the LMC.

Figure~\ref{fig:lf_bolo_witherrs} shows that the luminosity functions of SN progenitors and LMC RSGs are consistent with each other for magnitudes brighter than $-7\,\text{mag}$ and their $1\sigma$ regions overlap in all bins. We note that the uncertainty region appears to be large compared to the fluctuations between bins. The reason is that the bins are not independent of each other for a cumulative distribution. We conclude that the RSG luminosity function does not provide any indications for a red supergiant problem. This supports the result by \citet{davies2018, davies2020} who found that the RSG problem is not significant even when using the ordering method.

The two distributions in Fig.~\ref{fig:lf_bolo_witherrs} appear to have different shapes at $-8.5\,\text{mag}$. However, this is entirely due to the relatively low number of progenitor searches: The figure shows that $<1\%$ of the LMC RSGs are brighter than $-8.5\,\text{mag}$ which is consistent with finding zero as bright SN progenitors in $27$ sensitive searches (see Table~\ref{tab:rsg_progenitors}). If the progenitor distribution had a cutoff at a bolometric magnitude of $-8.25\,\text{mag}$, the discrepancy to the LMC sample would only become significant at the $2\sigma$ confidence level for more than ten times as many non-detections.

Even if there was a significant discrepancy between the luminosities of RSG SN progenitors and regular RSGs, it might be caused by differences between the two populations: Most LMC M supergiants are less evolved and hotter than the late-type bolometric correction that we adopted for RSG progenitor stars following \citet{davies2018}. 
Another difference is that the LMC is a low-extinction, low-metallicity environment, while SN progenitors are typically detected in massive galaxies that have been monitored with the \emph{Hubble Space telescope} \citep{davies2020}.
Therefore, we also compare the progenitor luminosity function to recently compiled M31 RSG samples. The sample by \citet{mcdonald2022_m31sample} is very similar to the LMC sample shown in Fig.~\ref{fig:lf_bolo_witherrs} and the sample by \citet{massey2021_rsgsample} is slightly lower around $-7\,\text{mag}$, but this discrepancy is less significant than $2\sigma$.

In addition, the impending core collapse might trigger changes at the surface of RSG progenitors: spectroscopy and photometry of young SNe indicate that a large fraction of RSGs undergo enhanced mass loss shortly before the SN explosion \citep{bruch2022, foerster2018_early_snlcs_csm, morazova2020}. A months-long stellar outburst was detected shortly before the explosion of one SN II with early interaction signatures \citep{jacobson-galan2022} and much brighter eruptions are frequent for interaction-powered SNe of Type IIn \citep{ofek2014, strotjohann2021}. Such outbursts could either increase the progenitor magnitude or they could dimm and redden the progenitor due to absorption or because the photosphere cools after expanding (see calculations by \citealt{davies2022, moriya2023_rsg_grid}). In summary, the very late stages of stellar evolution are not well understood and discrepancies to younger, less evolved RSGs would not necessarily imply the existence of failed SNe.

\section{Conclusion} \label{sec:conclusion}

Here, we calculate a bias-corrected luminosity function for RSG SN progenitors using limiting magnitudes to correct for the sensitivity of the search. Compared to previously presented luminosity functions, we observe fewer intermediate-luminosity progenitors while the fraction of faint progenitor stars increases, e.g., we find that only $36\pm11\%$ of the RSG SN progenitors are brighter than $-7\,\text{mag}$, compared to previously quoted fractions of $56\pm5\%$ \citep{davies2020}. The larger uncertainty on our estimate is due to statistical fluctuations that were so far only considered for the highest-luminosity progenitor \citep{davies2018}. Here, we generalize their approach and show that the statistical uncertainties are large for the entire luminosity function. Measurement errors, on the other hand, only have a minor impact as long as they fluctuate randomly for the different progenitors and are not systematically biased for a large fraction of the sample (see Sect.~\ref{sec:lf_with_errors}). The bias-corrected progenitor luminosity function is consistent with the luminosity functions for RSGs in the LMC and in M31.
So, while failed SNe may exist, our work shows that the RSG SN progenitor luminosity function does not motivate a large fraction of failed SNe.

Finally, we would like to highlight that the presented statistical method is not limited to the progenitor luminosity function. It applies to any study with a fixed sample size in which not every search is sensitive to the probed characteristic or quantity, which could, for example, be the morphology, the appearance of spectral or light curve features, or the presence of multiwavelength or multimessenger signals.

\begin{acknowledgments}

We would like to thank Morgan Fraser, Emma Beasor, Jonathan Mushkin, and Doron Kushnir for helpful discussion, and Barak Zackay for comments on the manuscript. We are grateful to the organizers of the MIAPbP workshop on interacting supernovae. Moreover, we would like to thank the referee for their constructive and helpfull feedback.

N.L.S. is funded by the Deutsche Forschungsgemeinschaft (DFG, German Research Foundation) via the Walter Benjamin program – 461903330. This research was supported by the Munich Institute for Astro-, Particle and BioPhysics (MIAPbP) which is funded by the Deutsche Forschungsgemeinschaft under Germany's Excellence Strategy – EXC-2094 – 390783311.

E.O.O. is grateful for the support of grants from the Benozio center, Willner Family Leadership Institute, Ilan Gluzman (Secaucus NJ), Madame Olga Klein - Astrachan, Minerva foundation, Israel Science Foundation, NSF, Israel Ministry of Science, Yeda-Sela, and Weizmann-MIT.

This research has made use of the NASA/IPAC Extragalactic Database (NED), which is funded by the National Aeronautics and Space Administration and operated by the California Institute of Technology.

AGY’s research is supported by the EU via ERC grant No. 725161, the ISF GW excellence center, an IMOS space infrastructure grant and a GIF grant, as well as the André Deloro Institute for Advanced Research in Space and Optics, The Helen Kimmel Center for Planetary Science, the Schwartz/Reisman Collaborative Science Program and the Norman E Alexander Family M Foundation ULTRASAT Data Center Fund, Minerva and Yeda-Sela;  AGY is the incumbent of the The Arlyn Imberman Professorial Chair.
\end{acknowledgments}

\bibliography{references}{}
\bibliographystyle{aasjournal}

\begin{longrotatetable}
\begin{deluxetable*}{clccccccccl}
\tablecaption{\label{tab:rsg_progenitors}Limits and detections of RSG progenitor stars from the literature.}
\tablehead{\colhead{\#} & \colhead{SN} & \colhead{dist. mod.} & \colhead{$E(B-V)$} & \colhead{band} & \colhead{BC} & \colhead{mag} & \colhead{limmag} & \colhead{abs. bol. mag} & \colhead{abs. bol. limmag} & \colhead{Ref.} \\
\colhead{}  & \colhead{}  & \colhead{(mag)} & \colhead{(mag)} & \colhead{} & \colhead{(mag)} & \colhead{(mag)} & \colhead{(mag)} & \colhead{(mag)} & \colhead{(mag)} & \colhead{}}
\startdata
1 & SN\,2017eaw & $ 29.44 \pm {\it 0.12} $ & $ 0.3 \pm {\it 0.04} $ & $ F606W $ & $ -1.83 \pm 0.38 $ & $ 26.4 \pm 0.05 $ & $ 28.55 \pm {\it 0.1} $ & $ -5.53 \pm 0.41 $ & $ -3.38 \pm 0.42 $ & 1, 2\\ 
2 & SN\,2003gd & $ 29.84 \pm 0.19 $ & $ 0.14 \pm 0.04 $ & $ F814W $ & $ 0.1 \pm 0.15 $ & $ 24.0 \pm 0.04 $ & $ 26.39 \pm {\it 0.1} $ & $ -5.97 \pm 0.26 $ & $ -3.58 \pm 0.27 $ & 3, 4\\ 
3 & SN\,2008bk & $ 27.96 \pm 0.13 $ & $ 0.08 \pm 0.03 $ & $ K $ & $ 3.0 \pm 0.18 $ & $ 18.39 \pm 0.03 $ & $ 21.09 \pm {\it 0.1} $ & $ -6.6 \pm 0.22 $ & $ -3.9 \pm 0.24 $ & 3, 4, 5, 6\\ 
4 & SN\,2020jfo & $ 30.81 \pm {\it 0.12} $ & $ 0.02 \pm {\it 0.04} $ & $ F814W $ & $ 0.0 \pm 0.15 $ & $ 25.02 \pm 0.07 $ & $ 26.8 \pm {\it 0.1} $ & $ -5.82 \pm 0.22 $ & $ -4.04 \pm 0.23 $ & 7\\ 
5 & SN\,2005cs & $ 29.62 \pm 0.12 $ & $ 0.16 \pm 0.02 $ & $ F814W $ & $ 0.0 \pm 0.15 $ & $ 23.62 \pm 0.07 $ & $ 25.4 \pm {\it 0.1} $ & $ -6.27 \pm 0.21 $ & $ -4.49 \pm 0.22 $ & 3, 4\\ 
6 & SN\,2018aoq & $ 31.3 \pm {\it 0.12} $ & $ 0.03 \pm {\it 0.04} $ & $ F814W $ & $ 0.0 \pm 0.15 $ & $ 23.91 \pm 0.02 $ & $ 26.81 \pm {\it 0.1} $ & $ -7.44 \pm 0.2 $ & $ -4.54 \pm 0.23 $ & 1\\ 
7 & SN\,2018zd & $ 30.61 \pm {\it 0.12} $ & $ 0.08 \pm {\it 0.04} $ & $ F814W $ & $ 0.0 \pm 0.15 $ & $ 25.13 \pm 0.15 $ & $ 26.09 \pm {\it 0.1} $ & $ -5.61 \pm 0.25 $ & $ -4.66 \pm 0.23 $ & 8, 9 \\ 
8 & SN\,2013ej & $ 29.8 \pm 0.11 $ & $ 0.14 \pm 0.02 $ & $ F814W $ & $ 0.4 \pm 0.2 $ & $ 22.65 \pm 0.05 $ & $ 24.8 \pm {\it 0.1} $ & $ -6.99 \pm 0.24 $ & $ -4.84 \pm 0.25 $ & 2, 3 \\ 
9 & SN\,2022acko & $ 31.74 \pm {\it 0.12} $ & $ 0.17 \pm {\it 0.04} $ & $ F814W $ & $ 0.0 \pm 0.15 $ & $ 25.61 \pm 0.09 $ & $ 27.12 \pm {\it 0.1} $ & $ -6.41 \pm 0.22 $ & $ -4.9 \pm 0.23 $ & 10\\ 
10 & SN\,2023ixf & $ 29.18 \pm 0.05 $ & $ 0.04 \pm {\it 0.04} $ & $ F814W $ & $ -2.28 \pm {\it 0.1} $ & $ 24.44 \pm 0.06 $ & $ 26.39 \pm {\it 0.1} $ & $ -7.09 \pm 0.14 $ & $ -5.14 \pm 0.16 $ & 10, 11\\
11 & SN\,2012A & $ 29.96 \pm 0.15 $ & $ 0.03 \pm {\it 0.04} $ & $ K $ & $ 3.0 \pm 0.18 $ & $ 20.29 \pm 0.13 $ & $ 21.4 \pm {\it 0.1} $ & $ -6.68 \pm 0.27 $ & $ -5.57 \pm 0.26 $ & 3, 8\\ 
12 & SN\,2019mhm & $ 30.81 \pm {\it 0.12} $ & $ 0.18 \pm {\it 0.04} $ & $ F814W $ & $ 0.0 \pm 0.15 $ & $ -$ & $ 24.97 \pm {\it 0.1} $ & $ - $ & $ -6.14 \pm 0.23 $ & 13\\ 
13 & SN\,2004A & $ 31.54 \pm 0.17 $ & $ 0.21 \pm 0.06 $ & $ F814W $ & $ 0.0 \pm 0.15 $ & $ 24.36 \pm 0.12 $ & $ 25.56 \pm {\it 0.1} $ & $ -7.52 \pm 0.28 $ & $ -6.32 \pm 0.27 $ & 3, 4, 14\\ 
14 & SN\,2006ov & $ 30.5 \pm 0.19 $ & $ 0.08 \pm 0.02 $ & $ F814W $ & $ 0.0 \pm 0.15 $ & $ -$ & $ 24.2 \pm {\it 0.1} $ & $ - $ & $ -6.43 \pm 0.26 $ & 3, 15\\ 
15 & SN\,2008cn & $ 32.61 \pm 0.1 $ & $ 0.33 \pm 0.04 $ & $ F814W $ & $ 0.0 \pm 0.15 $ & $ 25.13 \pm 0.09 $ & $ 26.64 \pm {\it 0.1} $ & $ -8.02 \pm 0.21 $ & $ -6.51 \pm 0.22 $ & 3, 16\\ 
16 & SN\,2012ec & $ 31.19 \pm 0.1 $ & $ 0.22 \pm 0.06 $ & $ F814W $ & $ 0.0 \pm 0.15 $ & $ 23.39 \pm 0.08 $ & $ 25.03 \pm {\it 0.1} $ & $ -8.16 \pm 0.22 $ & $ -6.52 \pm 0.23 $ & 3\\ 
17 & SN\,2004et & $ 29.43 \pm {\it 0.12} $ & $ 0.42 \pm 0.03 $ & $ Ij $ & $ 0.25 \pm 0.15 $ & $ 22.06 \pm 0.12 $ & $ 23.26 \pm {\it 0.1} $ & $ -7.78 \pm 0.23 $ & $ -6.58 \pm 0.22 $ & 1, 3, 5, 15 \\ 
18 & SN\,2006my & $ 31.74 \pm 0.12 $ & $ 0.49 \pm 0.26 $ & $ F814W $ & $ 0.0 \pm 0.15 $ & $ 24.86 \pm 0.13 $ & $ 25.97 \pm {\it 0.1} $ & $ -7.7 \pm 0.48 $ & $ -6.59 \pm 0.48 $ & 3, 4, 15\\ 
19 & SN\,2006bc & $ 30.84 \pm 0.18 $ & $ 0.21 \pm {\it 0.04} $ & $ F814W $ & $ 0.0 \pm 0.15 $ & $ -$ & $ 24.45 \pm {\it 0.1} $ & $ - $ & $ -6.73 \pm 0.26 $ & 3\\ 
20 & SN\,2021yja & $ 31.84 \pm {\it 0.12} $ & $ 0.10 \pm {\it 0.04} $ & $ F606W $ & $ -1.83 \pm 0.38 $ & $ -$ & $ 27.07 \pm {\it 0.1} $ & $ - $ & $ -6.83 \pm 0.42 $ & 17 \\ 
21 & SN\,2004dg & $ 31.51 \pm 0.13 $ & $ 0.24 \pm 0.03 $ & $ F814W $ & $ 0.0 \pm 0.15 $ & $ -$ & $ 25.0 \pm {\it 0.1} $ & $ - $ & $ -6.91 \pm 0.23 $ & 3\\ 
22 & SN\,2007aa & $ 31.56 \pm 0.13 $ & $ 0.03 \pm {\it 0.04} $ & $ F814W $ & $ 0.0 \pm 0.15 $ & $ -$ & $ 24.44 \pm {\it 0.1} $ & $ - $ & $ -7.17 \pm 0.23 $ & 3\\ 
23 & SN\,1999gi & $ 30.0 \pm 0.08 $ & $ 0.17 \pm 0.04 $ & $ F606W $ & $ -1.83 \pm 0.38 $ & $ -$ & $ 24.9 \pm {\it 0.1} $ & $ - $ & $ -7.3 \pm 0.41 $ & 3, 5\\ 
24 & SN\,2012aw & $ 29.96 \pm 0.02 $ & $ 0.43 \pm 0.05 $ & $ K $ & $ 3.0 \pm 0.18 $ & $ 19.56 \pm 0.29 $ & $ 19.8 \pm {\it 0.1} $ & $ -7.56 \pm 0.34 $ & $ -7.32 \pm 0.21 $ & 3, 18\\ 
25 & SN\,2001du & $ 31.31 \pm 0.07 $ & $ 0.16 \pm 0.08 $ & $ F814W $ & $ 0.0 \pm 0.15 $ & $ -$ & $ 24.25 \pm {\it 0.1} $ & $ - $ & $ -7.32 \pm 0.24 $ & 3, 5\\ 
26 & SN\,2018ivc & $ 30.02 \pm {\it 0.12} $ & $ 0.5 \pm {\it 0.04} $ & $ F606W $ & $ -1.83 \pm 0.38 $ & $ -$ & $ 25.4 \pm {\it 0.1} $ & $ - $ & $ -7.53 \pm 0.42 $ & 19\\ 
27 & SN\,1999br & $ 30.75 \pm 0.18 $ & $ 0.02 \pm 0.02 $ & $ F606W $ & $ -1.83 \pm 0.38 $ & $ -$ & $ 24.9 \pm {\it 0.1} $ & $ - $ & $ -7.71 \pm 0.43 $ & 3, 5\\ 
28 & SN\,1999em & $ 30.34 \pm 0.09 $ & $ 0.09 \pm 0.04 $ & $ Ic $ & $ -0.32 \pm 0.15 $ & $ -$ & $ 23.0 \pm {\it 0.1} $ & $ - $ & $ -7.79 \pm 0.21 $ & 3, 5\\ 
29 & SN\,1999an & $ 31.34 \pm 0.08 $ & $ 0.11 \pm 0.05 $ & $ F606W $ & $ -1.83 \pm 0.38 $ & $ -$ & $ 24.7 \pm {\it 0.1} $ & $ - $ & $ -8.7 \pm 0.41 $ & 3, 5 \\ 
30 & SN\,2002hh & $ 29.43 \pm {\it 0.12} $ & $ 1.36 \pm 0.05 $ & $ i $ & $ -0.49 \pm 0.15 $ & $ -$ & $ 22.8 \pm {\it 0.1} $ & $ - $ & $ -9.23 \pm 0.23 $ & 1, 3, 5, 20\\ 
31 & SN\,2020fqv & $ 31.19 \pm {\it 0.12} $ & $ 0.52 \pm {\it 0.04} $ & $ F606W $ & $ -1.83 \pm 0.38 $ & $ -$ & $ 24.88 \pm {\it 0.1} $ & $ - $ & $ -9.27 \pm 0.42 $ & 21 \\ 
\enddata
\tablecomments{Progenitor searches sorted by sensitivity (second last column). All magnitudes are given in the Vega system and we adopt the bolometric corrections by \citet{davies2018}. SNe that are not in their sample are sometimes lacking uncertainties on the distance modulus or extinction. In these cases, we use the median uncertainties of all other SNe, i.e., an uncertainty of $0.12\,\text{mag}$ on the distance modulus and an uncertainty of $0.04\,\text{mag}$ on $E(B-V)$. In addition, we assume an uncertainty of $0.1\,\text{mag}$ ($\sim10\%$) for all limiting magnitudes. For the bolometric correction of SN\,2023ixf, we adopt an adhoc uncertainty of $0.1\,\text{mag}$. Italic font indicates that we are using default uncertainties. \\
(1) \citet{davies2020}; (2) \citet{vandyk2019_sn2017eaw}; (3) \citet{davies2018}; (4) \citet{maund2014}; (5) \citet{smartt2009}; (6) \citet{vandyk2013_sn2008bk_progenitor_vanished}; (7) \citet{sollerman2021}; (8) \citet{vandyk2023_disapperance_of_progenitor_stars}; (9) \citet{hiramatsu2021_sn2018zd}; (10) \citet{vandyk2023_progenitor_2022acko}; (11) \citet{kilpatrick2023_sn2023ixf_progenitor}; (12) \citet{jacobson-galan2023_sn2023ixf}; (13) \citet{vazquez2022_sn2019mjm_progenitor_nondet}; (14) \citet{maund2017_progenitor_stellar_environment}; (15) \citet{crockett2011}; 
(16) \citet{maund2015}; (17) \citet{hosseinzadeh2022_2021yja}; (18) \citet{fraser2016_disappearanceSN2012aw}; (19) \citet{bostroem2020}; (20) \citet{kochanek2020}; (21) \citet{tinyanont2022_sn2020fqv} \\
We exclude the less secure progenitor detections of SN\,2009hd (astrometry not precise enough; \citealt{elias-rosa2011}), SN\,2009kr (likely a stellar cluster; \citealt{maund2015}), SN\,2009md (still present in late-time observations; \citealt{maund2015}), and SN\,2016cok (not a single star; \citealt{kochanek2017_progenitor_sn2016cok}) from the sample.}
\end{deluxetable*}
\end{longrotatetable}

\end{document}